\documentclass[11pt]{article}
\usepackage{graphicx}
\usepackage{amsmath,amssymb,amsthm,amsfonts}
\usepackage{amssymb}
\newtheorem{thm}{Theorem}[section]

\newtheorem{lem}[thm]{Lemma}
\newtheorem{prop}[thm]{Proposition}

\theoremstyle{definition}

\theoremstyle{remark}

\numberwithin{equation}{section}

\DeclareMathSymbol{\C}{\mathalpha}{AMSb}{"43}

\newcommand{\eps}{\varepsilon}

\newcommand{\lam}{\lambda}
\newcommand{\alp}{\alpha}

\newcommand{\R}{{\mathbb{R}}}

\newcommand{\inte}{\int_{\mathbb{R}^3}}

\newcommand{\bsub}{\begin{subequations}}
\newcommand{\esub}{\end{subequations}$\!$}

\begin{document}
\title{The Lieb-Yau Conjecture for Ground States of Pseudo-Relativistic Boson Stars}

\author{Yujin Guo\thanks{Email: \texttt{yguo@mail.ccnu.edu.cn}.}
 \, and\, Xiaoyu Zeng\thanks{Email: \texttt{xyzeng@whut.edu.cn}.}
}

\date{\today}

\smallbreak \maketitle

\begin{abstract}
It is known that ground states of the pseudo-relativistic Boson stars exist if and only if the stellar mass $N>0$ satisfies $N<N^*$, where the finite constant $N^*$ is called the critical stellar mass. Lieb and Yau conjecture in [Comm. Math. Phys., 1987] that ground states of the pseudo-relativistic Boson stars are {\em unique} for each $N<N^*$. In this paper, we prove that the above uniqueness conjecture holds for the particular case where $N>0$ is small enough.
\end{abstract}

\vskip 0.2truein


\vskip 0.2truein

\section{Introduction}
Various models of pseudo-relativistic boson stars have attracted a
lot of attention in theoretical and numerical astrophysics over the past few decades, see \cite{LY,LY87}
and references therein.
In this paper, we are interested in ground states of pseudo-relativistic Boson stars in the mean field limit (cf. \cite{ES,FJCMP,LY}), which can be described by constraint minimizers of the following variational problem
\begin{equation}\label{def:eN}
e(N):=\inf \Big\{  \mathcal{E}(u):\, u\in H^{\frac{1}{2}}(\R^3)\,\ \mbox{and}\ \int_{\R ^3} |u(x)|^2dx=N\Big\},
\end{equation}
where $N>0$ denotes the stellar mass of Boson stars, and the pseudo-relativistic Hartree energy functional $ \mathcal{E}(u)$ is of the form
\begin{equation}\label{f}
  \mathcal{E}(u):=\int_{\R ^3} \bar u\big( \sqrt{-\Delta +m^2}-m\big)udx-\frac{1}{2}\int_{\R ^3}\big(|x|^{-1}\ast |u|^2\big)|u|^2dx,\ \ m>0.
\end{equation}
Here the operator $\sqrt{-\Delta +m^2}$ is defined via multiplication in the Fourier space with the symbol $\sqrt{|\xi|^2+m^2}$ for $\xi\in\R^3$, which describes the kinetic and rest energy of many self-gravitating and relativistic bosons with rest mass $m>0$, and the symbol $\ast$ stands for the convolution on $\R^3$. Because of the physical relevance, without special notations we always focus on the case  $m>0$ throughout the whole paper. The main purpose of this paper is to prove the uniqueness of minimizers for $e(N)$, provided that $N>0$ is small enough.

The variational problem $e(N)$ is essentially in the class of $L^2-$critical constraint minimization problems, which were studied recently in the nonrelativistic cases, e.g. \cite{Cao,GLW,GS,GZZ} and references therein.
Comparing with these mentioned works, it however deserves to remark that the analysis of $e(N)$ is more complicated in a substantial way, which is mainly due to the nonlocal nature of the pseudo-differential operator $\sqrt{-\Delta +m^2}$, and the convolution-type nonlinearity as well. Starting from the pioneering papers \cite{LY,LY87}, many works were devoted to the mathematical analysis of the variational problem $e(N)$ over the past few years, see \cite{ES,FL,FJ,FJCMP,FrL,GZ,L09,N,YY} and references therein.
The existing results show that the analysis of $e(N)$ is connected well to the following Gagliardo-Nirenberg inequality  of fractional type
\begin{equation}\label{GNineq}
\inte \big(\frac{1}{|x|}\ast |u|^2\big)|u|^2 dx\le \frac 2 {\|w\|^2_2}\|(-\Delta )^{1/4}u\|^2_2\, \|u\|^2_2 ,\   \  u \in H^{\frac{1}{2}}(\R^3),
\end{equation}
where $w(x)=w(|x|)>0$ is a ground state, up to translations and suitable rescaling (cf. \cite{FJCMP,LY}), of the fractional equation
\begin{equation}
\sqrt{-\Delta}\,u+u-\Big(\frac{1}{|x|}\ast |u|^2\Big)u=0
\  \mbox{  in } \  \R^3,\  \mbox{ where }\  u\in H^{\frac{1}{2}}(\R ^3).
\label{w:eqn}
\end{equation}
By making full use of (\ref{GNineq}), Lenzmann in \cite{L09} established the following interesting existence and analytical characters of minimizers for $e(N)$:

\vskip 0.05truein

\noindent {\bf Theorem A (\cite[Theorem 1]{L09})} {\em
Under the assumption $m>0$, the following results hold for $e(N)$:
\begin{enumerate}
\item  $e(N)$ has minimizers if and only if $0<N<N^*$, where the finite constant $N^*$ is independent of $m$.

\item Any minimizer $u$ of $e(N)$ satisfies $u\in H^s(\R^3)$ for all $s\ge \frac{1}{2}$.
\item Any nonnegative minimizer of $e(N)$ must be strictly positive and radially symmetric-decreasing, up to phase and translation.
\end{enumerate}
}

We remark that the existence of the critical constant $N^*>0$ stated in Theorem A, which is called the critical stellar mass of boson stars, was proved earlier in \cite{FJCMP,LY}. Further, the dynamics and some other analytic properties of minimizers for $e(N)$ were also investigated by Lenzmann and his collaborators in \cite{FL,FJ,FJCMP,FrL,L,L09} and references therein. Stimulated by \cite{GS,GZZ}, the related limit behavior of minimizers for $e(N)$ as $N\nearrow N^*$ were studied more recently in \cite{GZ,N,YY}, where the Gagliardo-Nirenberg type inequality (\ref{GNineq}) also played an important role. Note also from the variational theory that any minimizer $u$ of $e(N)$ satisfies the Euler-Lagrange equation
\begin{equation}\label{1:eq1.18}
\big( \sqrt{-\Delta +m^2}-m\big)u- \Big(\frac{1}{|x|}\ast |u|^2\Big)u =-\mu  u  \ \  \text{in}\, \
\mathbb{R}^3,
\end{equation}
where $\mu \in \R$ is the associated Lagrange multiplier. Following (\ref{1:eq1.18}) and Theorem A, one can deduce that any minimizer of $e(N)$ must be either positive or negative, see \cite{GZ} for details. Therefore, it is enough to consider positive minimizers of $e(N)$, which are called ground states of $e(N)$ throughout the rest part of this paper.

Whether a physics system admits a unique ground state or not is an interesting and fundamental problem.
Lieb and Yau  \cite{LY} conjectured in 1987 that for each $N<N^*$, there exists a unique ground state (minimizer) of $e(N)$. As expected by Lieb and Yau there, the analysis of this uniqueness conjecture is however challenging extremely. Actually, it is generally difficult to prove whether any two different ground states of $e(N)$ satisfy the equation (\ref{1:eq1.18}) with {\em the same} Lagrange multiplier $\mu \in \R$. On the other hand, more difficultly, it seems very challenging to address the uniqueness of ground states for (\ref{1:eq1.18}). Essentially, even though the uniqueness of ground states for the following fractional equation
 \begin{equation}
(-\Delta)^s\,u+u-u^{\alpha +1}=0
\  \mbox{  in } \  \R^N,\     u\in H^s(\R ^N) ,
\label{p:eqn}
\end{equation}
where $0<s<1$, $\alpha >0$ and $N\ge 1$,
was already proved in the celebrated works \cite{FL13,FL}, the uniqueness of ground states for (\ref{w:eqn}) or  (\ref{1:eq1.18}) is still open, due to the nonlocal nonlinearity of Hartree type. Therefore, whether the above Lieb-Yau conjecture is true for all $0<N<N^*$ remains mainly open after three decades, except Lenzmann's  recent work \cite{L09} in 2009.

As an important first step towards the Lieb-Yau conjecture, Lenzmann proved in \cite{L09} that  for each $0<N\ll N^*$ and except for at most countably many $N$, the uniqueness of minimizers for $e(N)$ holds true. We emphasize that the additional assumption ``except for at most countably many $N$" seems essential in Lenzmann's proof, since the smoothness of the GP energy $e(N)$ with respect to $N$ was employed there.
In this paper we intend to remove the above additional assumption and prove the Lieb-Yau conjecture in the particular case where $0<N<N^*$ is small enough. More precisely, our main result of this paper is the following uniqueness of minimizers for $e(N)$.

\begin{thm}\label{th1}
If $N> 0$ is small enough, then the problem $e(N)$ admits a unique positive minimizer, up to phase and translation.
\end{thm}

The similar uniqueness results of Theorem \ref{th1} were established recently in \cite[Theorem 2]{AF} and \cite[Theorem 1.1]{M} (see also \cite[Corollary 1.1]{GWZZ}) for the nonrelativistic Hartree minimization problems with trapping potentials, under the additional assumption that the associated nonrelativistic operator admits the first eigenvalue. We however emphasize that the arguments of \cite{AF,M,GWZZ} are not applicable for proving Theorem \ref{th1}, since the associated pseudo-relativistic operator $H:= \sqrt{-\Delta +m^2}-m$ does not admit any eigenvalue in our problem $e(N)$. Therefore, a different approach is needed for proving  Theorem \ref{th1}. Towards this purpose, since positive minimizers $u$ of $e(N)$ vanish uniformly as $N\searrow 0$, motivated by \cite{L09} we define
$$\tilde u(x)=c^2u(cx),\ \ \mbox{where}\ \ c>0,$$
so that
\begin{equation}\label{G2:1}
\mathcal{E}(u)=c^{-3}\mathcal{E}_c(\tilde u) \ \text{ and }\inte |\tilde u|^2dx=c\inte | u|^2dx,
\end{equation}
where the energy functional $\mathcal{E}_c(\cdot)$ is given by
\begin{equation}
\mathcal{E}_c(u):=\int_{\R ^3} \bar{ u}\big( \sqrt{-c^2\Delta +c^4m^2}-c^2m\big) udx-\frac{1}{2}\int_{\R ^3}\big(|x|^{-1}\ast |u|^2\big)|u|^2dx,\ \ m>0.
\end{equation}
Consider the minimization problem
\begin{equation}\label{def:ec}
\bar e(c):=\inf \Big\{  \mathcal{E}_c(u):\, u\in H^{\frac{1}{2}}(\R^3)\,\ \mbox{and}\ \int_{\R ^3} |u(x)|^2dx=1\Big\}.
\end{equation}
Note from Theorem A and (\ref{G2:1}) that if $c>0$ is large enough, then $\bar e(c)$ in (\ref{def:ec}) admits positive  minimizers.
More importantly, setting $c=N^{-1}>0$, studying positive minimizers of $e(N)$ as $N\searrow 0$ is then equivalent to investigating positive minimizers of the   minimization problem (\ref{def:ec}) as $c\nearrow \infty$.
Therefore, to establish the uniqueness of Theorem \ref{th1}, it suffices  to prove the following uniqueness theorem.

\begin{thm}\label{th2}
If $c> 0$ is large enough, then $\bar e(c)$ in (\ref{def:ec}) admits a unique positive minimizer, up to phase and translation.
\end{thm}

Suppose now that $Q_c>0$ is a minimizer of $\bar e(c)$ defined in (\ref{def:ec}). Then there exists a Lagrange multiplier $\mu_c \in \R$ such that $Q_c>0$ solves
 \begin{equation}\label{1D:eq1.18}
\big( \sqrt{-c^2\Delta +c^4m^2}-c^2m\big)Q_c- \Big(\frac{1}{|x|}\ast |Q_c|^2\Big)Q_c =-\mu_c  Q_c  \ \  \text{in}\, \
\mathbb{R}^3.
\end{equation}
Recall from \cite[Proposition 1]{L09} that up to a subsequence if necessary,  $\mu_c\in\R$   satisfies
\begin{equation}\label{1E:eq1.18} \mu_c\to -\lam <0 \ \ \mbox{as}\ \  c\to \infty
\end{equation}
for some constant $\lam >0$.
In order to prove Theorem \ref{th2}, associated to $Q_c$, we need to study the uniformly exponential decay of $\phi_c\in H^{\frac{1}{2}}(\R^3)$ as $c\to\infty$, where $\phi_c$ satisfies
\begin{equation}\label{1:e:1}
\begin{split}
 &\big(\sqrt{-c^2\Delta +m^2c^4}-mc^2\big)\phi_c -\Big(\frac{1}{|x|}\ast Q_c^2\Big)\phi_c  -2k_1\Big\{\frac{1}{|x|}\ast \big(Q_c \phi_c \big)\Big\}Q_c\\
 &-k_2(c)Q_c=-\mu_c \phi_c \ \text{ in}\ \ \R^3.
\end{split}
\end{equation}
Here $\mu_c\in\R$ satisfies (\ref{1E:eq1.18}),  $k_1\ge 0$ and $k_2(c)\in \R$ is bounded uniformly in $c>0$. As proved in Lemma \ref{lem:2.1}, we shall derive the following uniformly exponential decay of $\phi_c$ as $c\to\infty$:
\begin{equation}\label{D:eq1.09}
  |\phi_{c}(x)| \le Ce^{-\delta|x|} \ \ \mbox{in}\ \, \R^3
\end{equation}
holds uniformly as $c\to\infty$, where the constants $\delta>0$ and $C=C(\delta)>0$ are independent of $c$.  Since $\mu_c\in\R$   satisfies (\ref{1E:eq1.18}), stimulated by \cite{AS,FJCMP,H,SW}, the proof of (\ref{D:eq1.09}) is based on the uniformly exponential decay (\ref{E:e:3}) of the Green's function $G_c(\cdot)$ for  $\big(\bar H_c+\mu _c\big)^{-1}$ as $c\to\infty$, where the operator $\bar H_c$ is defined by
$$\bar H_c:=\sqrt{-c^2\Delta +m^2c^4}-mc^2.$$
As shown in Lemma \ref{lem:2.3}, it however deserves to remark that because $G_c(\cdot)$ depends on $c>0$, one needs to carry out more delicate analysis, together with some tricks, for addressing the  uniformly exponential decay (\ref{E:e:3}) of $G_c(\cdot)$ as $c\to\infty$. On the other hand, as a byproduct, the exponential decay (\ref{D:eq1.09}) can be useful in analyzing the limiting procedure of solutions for Schrodinger equations involving the above fractional operator $\bar H_c$, which were investigated widely in \cite{CLM,Choi,CN}
and references therein.

Following (\ref{D:eq1.09}) and the regularity of $Q_c$, in Section 2 we shall finally prove the following limit behavior
\begin{equation}\label{D:1.14}
Q_c\to Q_\infty  \ \mbox{ uniformly in }\  \R^3\  \text{ as }\ c \to \infty,
\end{equation}
where $Q_\infty>0$ is the unique positive minimizer of (\ref{def:em}) described below. Based on the refined estimates of Section 2, motivated by \cite{Deng,Grossi,GLW}, we shall employ the nondegenerancy and uniqueness of $Q_\infty$ to complete the proof of Theorem \ref{th2} by establishing Pohozaev identities.

This paper is organized as follows. In Section 2 we shall address some refined estimates of $Q_c$ as $c\to\infty$, where Lemma \ref{lem:2.1} is proved in Subsection 2.1. Following those estimates of Section 2, Section 3 is devoted to the proof of Theorems \ref{th2} on the uniqueness of minimizers for $\bar e(c)$ as $c\to\infty$. Theorem \ref{th1} then follows immediately from Theorem \ref{th2} in view of the relation (\ref{G2:1}).

\section{Analytical Properties of $Q_c$ as $c\to\infty$}

The main purpose of this section is to give some refined analytical estimates of $Q_c$ as $c\to\infty$, where $Q_c>0$ is a positive minimizer of $\bar e(c)$ defined in (\ref{def:ec}). Note also from Theorem A and (\ref{G2:1}) that $Q_c$ is radially symmetric in $|x|$.

We first introduce the following limit problem associated to $\bar e(c)$:
\begin{equation}\label{def:em}
{ e}_m:=\inf \Big\{  {E}_m(u):\, u\in H^{1}(\R^3)\,\ \mbox{and}\ \int_{\R ^3} |u(x)|^2dx=1\Big\},
\end{equation}
where the energy functional $E_m(\cdot)$ satisfies
\begin{equation}
{E}_m(u):=\frac{1}{2m}\int_{\R ^3} |\nabla u|^2dx-\frac{1}{2}\int_{\R ^3}\big(|x|^{-1}\ast |u|^2\big)|u|^2dx, \ \forall u\in H^1(\R^3),\ \ m>0.
\end{equation}
For any given $m>0$, it is well-known that, up to translations,  problem (\ref{def:em}) has a unique positive minimizer denoted by $Q_\infty(|x|)$, which must be radially symmetric, see \cite{L09,L76} and references therein. Further,  $Q_\infty>0$ solves the following equation
\begin{equation}\label{eq1.8}
-\frac{1}{2m}\Delta Q_\infty+\lambda Q_\infty=(|x|^{-1}\ast |Q_\infty|^2)Q_\infty\ \ \mbox{in}\, \ \R^3,\ \ Q_\infty\in H^1(\R^3),
\end{equation}
where the Lagrange multiplier  $\lambda >0$ depends only on $m$ and is determined uniquely by the constraint condition $\int_{\R ^3} Q_\infty ^2dx=1$. Note from \cite[Theorem 3]{MS} that $Q_\infty>0$ is a unique positive solution of (\ref{eq1.8}). Moreover, recall from \cite[Theorem 4]{L09} that $Q_\infty$ is non-degenerate, in the sense that the linearized operator $L_+: \, H^2(\R^3) \mapsto L^2(\R^3)$, which is defined by
\begin{equation}\label{eq1.10}
  L_+\xi:=\Big(-\frac{1}{2m}\Delta+\lambda-|x|^{-1}\ast |Q_\infty|^2\Big)\xi-2\big(|x|^{-1}\ast(Q_\infty\xi)\big)Q_\infty,
  \end{equation}
satisfies
\begin{equation}\label{eq1.12}
\ker L_+=\text{\rm span} \Big\{\frac{\partial Q_\infty}{\partial x_1}, \frac{\partial Q_\infty}{\partial x_2}, \frac{\partial Q_\infty}{\partial x_3}\Big\}.
\end{equation}

As a positive minimizer of $\bar e(c)$, $Q_c>0$ satisfies the following equation
\begin{equation}\label{eq2.7}
\big(\sqrt{-c^2\Delta +m^2c^4}-mc^2\big)Q_c-(|x|^{-1}\ast Q_c^2)Q_c=\mu_c Q_c \ \text{ in}\,\ \R^3,
\end{equation}
where $\mu_c\in \R$ is a suitable  Lagrange multiplier. Recall from \cite[Proposition 1]{L09} that up to a subsequence if necessary, the Lagrange multiplier  $\mu_c$ of (\ref{eq2.7}) satisfies
\begin{equation}\label{eq1.09C}
  \mu_c\to -\lam <0 \ \, \text{ as }\ c\to \infty,
\end{equation}
where $\lam>0$ is the same as that of (\ref{eq1.8}). Associated to the positive minimizer $Q_c>0$ of (\ref{eq2.7}), we next define the linearized operator
\begin{equation}\label{e:1}
\begin{split}
\mathcal{L}_{k_1,k_2}\xi:=&\big(\sqrt{-c^2\Delta +m^2c^4}-mc^2\big)\xi-\Big(\frac{1}{|x|}\ast Q_c^2\Big)\xi\\
&-2k_1\Big\{\frac{1}{|x|}\ast \big(Q_c \xi\big)\Big\}Q_c-k_2(c)Q_c \ \text{ in}\ \ \R^3,
\end{split}
\end{equation}
where the constants $m>0$ and $k_1\ge 0$, and $k_2(c)\in \R$ is bounded uniformly in $c>0$.

\begin{lem}\label{lem:2.1}
Suppose $\varphi_c=\varphi_c(x)\in H^{\frac{1}{2}}(\R^3)$ is a solution of
\begin{equation}\label{e:2}
\mathcal{L}_{k_1,k_2}\varphi_c=\mu_c \varphi_c \ \text{ in}\,\ \R^3,
\end{equation}
where $\mu_c$ satisfies (\ref{eq1.09C}), and the operator $\mathcal{L}_{k_1,k_2}$ is defined by (\ref{e:1}) for some constants $m>0$ and $k_1\ge 0$, and $k_2(c)\in \R$ being bounded uniformly in $c>0$. Then there exist $\delta>0$ and $C=C(\delta)>0$, which are independent of $c>0$, such that
 \begin{equation}\label{e:3}
|\varphi_c(x)|\le Ce^{-\delta |x|}\ \text{ in}\,\ \R^3
\end{equation}
uniformly for all sufficiently large $c>0$.
\end{lem}

Since the proof of Lemma \ref{lem:2.1} is a little involved, we leave it to Subsection 2.1. Applying Lemma \ref{lem:2.1}, we next address the following estimates of $Q_c>0$ as $c\to \infty$, which are crucial for the proof of Theorem \ref{th2}.

\begin{prop}
Let $Q_{c}>0$ be a positive  minimizer of $\bar e(c)$ defined in (\ref{def:ec}) as $c\to\infty$. Then we have
\begin{enumerate}
\item  There exist $\delta>0$ and $C=C(\delta)>0$, which are independent of $c>0$,  such that
  \begin{equation}\label{eq1.09}
  |Q_{c}(x)|,\, |\nabla Q_{c}(x)|\le Ce^{-\delta|x|} \ \ \mbox{in}\ \, \R^3
  \end{equation}
  uniformly as $c\to\infty$.

\item  $Q_{c}$ satisfies
  \begin{equation}\label{eq1.09B}
  Q_{c}\to Q_\infty  \ \mbox{ uniformly in }\  \R^3\  \text{ as }\ c\to \infty,
  \end{equation}
  where $Q_\infty>0$ is the unique positive  minimizer of (\ref{def:em}).
\end{enumerate}
\end{prop}

\noindent \textbf{Proof.} 1. Since $Q_{c}>0$ solves (\ref{eq2.7}), the uniformly exponential decay (\ref{eq1.09}) for $Q_{c}$ as $c\to\infty$ follows directly from (\ref{eF:3}) below. Because $\frac{\partial Q_{c}}{\partial x_i}$ satisfies (\ref{e:2}) for $k_1=1$ and $k_2(c)\equiv 0$, where $i=1, 2, 3$, the exponential decay (\ref{eq1.09}) holds for $\nabla Q_{c}$ by applying Lemma \ref{lem:2.1}.

2. Following \cite[Lemma 4.9]{M} and references therein, we first recall from \cite[Proposition 1]{L09} that $Q_{c}$ satisfies
\begin{equation}\label{eq1.9}
  Q_{c}\to Q_\infty  \ \text{ in } H^1(\R^3)\  \text{ as } c\to \infty,
\end{equation}
where the convergence holds for the whole sequence of $\{c\}$, due to the uniqueness of $Q_\infty>0$.
Rewrite (\ref{eq2.7}) as
\begin{equation}\label{2A:2}
 \begin{split}
\Big(-\frac{1}{2m}\Delta -\mu_{c}\Big)Q_{c}&=\big(|x|^{-1}\ast Q_{c}^2\big)Q_{c}-   \Big(\sqrt{-c^2\Delta +m^2c^4}-mc^2+\frac{1}{2m}\Delta \Big)Q_{c}\\
&:= \big(|x|^{-1}\ast Q_{c}^2\big)Q_{c}- F_{c} (\nabla)Q_{c}   \ \text{ in}\,\ \R^3,
\end{split}
\end{equation}
where we denote the pseudo-differential operator
\begin{equation}\label{2D:2}
F_{c}(\nabla):=\sqrt{-c^2\Delta +m^2c^4}-mc^2-\frac{-\Delta}{2m}
\end{equation}
with the symbol
\[
F_{c}(\xi):=\sqrt{c^2|\xi|^2 +m^2c^4}-mc^2-\frac{ |\xi|^2}{2m} ,\ \ \xi\in\R^3.
\]

Recall from (\ref{eq1.09}) that $Q_{c}$ decays exponentially as $|x|\to\infty$ for all sufficiently large $c>0$. Moreover, since the operator $\bar H_c:=\sqrt{-c^2\Delta +m^2c^4}-mc^2$ is uniformly bounded from below for all $c>0$,
the similar argument of \cite[Theorem 4.1(i)]{FJCMP} or \cite[Proposition 4.2]{Choi} applied to (\ref{eq2.7}) and (\ref{eq1.09C}) yields that
\begin{equation}\label{B:M}
Q_{c}\in H^s(\R^3) \ \   \mbox{for all}\, \ s\ge \frac{1}{2},
\end{equation}
which further implies the uniform smoothness of $Q_c$ in $c>0$, and
\begin{equation*}\label{2A:5OK}
Q_{c}\in L^\infty(\R^3) \ \   \mbox{uniformly in}\, \ c>0.
\end{equation*}
Applying the Taylor expansion, we obtain from (\ref{2D:2}) that for $|\xi|\ge \frac{mc}{2}$,
\[
\big|F_{c}(\xi)\big|=\Big|c|\xi|\sqrt{ 1+\big(\frac{mc}{|\xi|}\big)^2}-mc^2-\frac{ |\xi|^2}{2m} \Big|\le  \sqrt 5\, c|\xi|+mc^2+\frac{ |\xi|^2}{2m} \le \frac{36 |\xi|^4}{m^3c^2},\ \ \xi\in\R^3,
\]
and for $|\xi|\le \frac{mc}{2}$,
\[
\big|F_{c}(\xi)\big|=\Big|mc^2\sqrt{ 1+\big(\frac{|\xi|}{mc}\big)^2}-mc^2-\frac{ |\xi|^2}{2m} \Big|\le \frac{|\xi|^4}{8m^3c^2},\ \ \xi\in\R^3,
\]
due to the fact that $|\sqrt{1+t}-1-\frac{1}{2}t|\le\frac{1}{8}t^2$ holds for all $0<t\le \frac{1}{2}$. Following above estimates, we then derive from (\ref{2D:2}) and (\ref{B:M}) that for sufficiently large $c>0$,
\begin{equation}\label{2D:2D}
\|F_{c} (\nabla)Q_{c}\|_{H^s(\R^3)}\le \frac{M_1}{m^3c^2}\|Q_c\|_{H^{s+4}(\R^3)}<\frac{M_2}{m^3c^2} \ \ \mbox{for all}\ \ s\ge \frac{1}{2},
\end{equation}
where $M_1>0$ and $M_2>0$ are independent of $c>0$. Also, since
\begin{equation}\label{2A:E1}
\big\||x|^{-1}\ast Q_{c}^2\big\|_{L^\infty(\R^3)}\le C\big(\big\|Q_{c}\big\|^2_{L^4(\R^3)}+\big\|Q_{c}\big\|^2_{L^2(\R^3)}\big),
\end{equation}
we derive from  (\ref{B:M}) that for any $p\ge 2$,
\begin{equation}\label{2A:1M}
  \big\|\big(|x|^{-1}\ast Q_{c}^2\big)Q_{c}\big\|_{L^p(\R^3)}\le C\big(\big\|Q_{c}\big\|^2_{L^4(\R^3)}+\big\|Q_{c}\big\|^2_{L^2(\R^3)}\big)\big\|Q_{c}\big\|_{L^p(\R^3)}\le K_p,
\end{equation}
where $K_p>0$ is independent of $c>0$.  Employing (\ref{2D:2D}) and (\ref{2A:1M}), together with Sobolev imbedding theorem, the bootstrap argument applied to (\ref{2A:2}) yields that
\begin{equation}\label{eq2.20}
\text{$Q_{c}\to Q_\infty$ uniformly on any compact domain of $\R^3$ as $c\to\infty$.}
\end{equation}

On the other hand, one can easily deduce from (\ref{eq1.8}) that $Q_\infty$  decays exponentially as $|x|\to\infty$. Together with (\ref{eq1.09}), this indicates that for any $\eps>0$, there exists a constant $R_\eps>0$, independent of $c>0$, such that
$$|Q_c(x)|,\,|Q_\infty(x)|<\frac{\eps}{4}\ \, \text{ for any }\ |x|>R_\eps,$$
and hence,
$$\sup_{|x|>R_\eps}|Q_c(x)-Q_\infty(x)|\leq\sup_{|x|>R_\eps} (|Q_c(x)|+|Q_\infty(x)|)\leq \frac{\eps}{2}.$$
Moreover, it follows from (\ref{eq2.20}) that for sufficiently large  $c>0$,
$$\sup_{|x|\leq R_\eps}|Q_c(x)-Q_\infty(x)|\leq\frac{\eps}{2}.$$
The above two estimates thus yield that for sufficiently large  $c>0$,
$$\sup_{x\in\R^3}|Q_c(x)-Q_\infty(x)|\leq\eps,$$
which implies that (\ref{eq1.09B}) holds true. The lemma is therefore proved.
\qed

\subsection{Uniformly exponential decay as $c\to\infty$}

In this Subsection, we address the proof of Lemma  \ref{lem:2.1} on the uniformly exponential decay as $c\to\infty$. We remark that even though the proof of Lemma  \ref{lem:2.1} is stimulated from \cite{AS,FJCMP,H,SW},  as shown in proving Lemma \ref{lem:2.3} below, we need to carry out more delicate analysis together with some tricks.

We first suppose that $\varphi_c\in H^{\frac{1}{2}}(\R^3)$ is a  solution of
\begin{equation}\label{R:e:2}
\big(\sqrt{-c^2\Delta +m^2c^4}-mc^2\big)\varphi_c-\Big(\frac{1}{|x|}\ast \varphi_c^2\Big)\varphi_c=-\lam _c \varphi_c \ \text{ in}\,\ \R^3,
\end{equation}
where the constant $m>0$ and
\begin{equation}\label{R:e:2M}
  \lam _c\to 2\lam >0 \ \, \text{ as }\ c\to \infty
\end{equation}
for some positive constant $\lam $. Define
\begin{equation}\label{R:e:2D}
\bar H_c:=\sqrt{-c^2\Delta +m^2c^4}-mc^2,\quad V(x):=-\frac{1}{|x|}\ast \varphi_c^2.
\end{equation}
Therefore, $\varphi_c$ can be thought of as an eigenfunction of the Schrodinger operator $H:=\bar H_c+V(x)$. Moreover, the argument of \cite[Theorem 4.1]{FJCMP} or \cite[Proposition 4.2]{Choi} gives that $\varphi_c\in H^s(\R^3)$ for all $s\ge \frac{1}{2}$, which implies the smoothness of $\varphi_c$. Further, the spectrum of $\bar H_c$ satisfies
\[
\sigma (\bar H_c)=\sigma _{ess}(\bar H_c)=[0,\infty)
\]
for all $c>0$. Under the assumption (\ref{R:e:2M}), then  $\big(\bar H_c+\lam _c \big)^{-1}$ exists for all sufficiently large $c>0$, and (\ref{R:e:2}) can be rewritten as
\[
\varphi_c(x)=-\big(\bar H_c+ \lam _c\big)^{-1}V(x)\varphi_c(x).
\]
Note also from (\ref{R:e:2}) and (\ref{R:e:2D}) that
 \begin{equation}\label{R:e:4}
\varphi_c(x)=-\inte G_c(x-y)V(y)\varphi_c(y)dy,
\end{equation}
where $G_c(x-y)$ is the  Green's function of $\big(\bar H_c+\lam _c\big)^{-1}$ defined in (\ref{R:e:2D}). The following lemma gives the uniformly exponential decay of  $G_c(\cdot)$ as $c\to\infty$.

\begin{lem}\label{lem:2.3} Suppose $\lam _c\in \R$ satisfies (\ref{R:e:2M}) for some $\lam >0$. Then for each $0<\delta < \min\{ \frac{m}{2},\sqrt{\lam m}\}$, there exists a constant $M:=M(\delta)>0$, independent of $c>0$, such that the Green's function $G_c(x-y)$ of $\big(\bar H_c+\lam _c\big)^{-1}$ satisfies
 \begin{equation}\label{E:e:3}
|G_c(x-y)|\le M\frac{e^{-\delta |x-y|}}{|x-y|^2}\ \text{ in}\,\ \R^3
\end{equation}
uniformly for all sufficiently large $c>0$.
\end{lem}

\noindent \textbf{Proof.}  Under the assumption (\ref{R:e:2M}), since $G_c(\cdot)$ is the  Green's function of $\big(\bar H_c+\lam _c\big)^{-1}$ defined in (\ref{R:e:2D}) for all sufficiently large $c>0$, we have
 \begin{equation}\label{R:e:5}
G_c(z)=f_c^{-1}(z), \ \, f_c(\mu )=\frac{1}{\sqrt{c^2|\mu|^2 +m^2c^4}-mc^2+\lam _c}\quad \mbox{for} \ \ \mu\in \R^3,
\end{equation}
and $f_c^{-1}: \mathcal{S'}\to \mathcal{S}'$ denotes the inverse Fourier transform of $f_c$. We obtain from (\ref{R:e:5}) that for all sufficiently large $c>0$,
 \begin{equation}\label{R:e:6}
G_c(z)=f_c^{-1}(z)=(2\pi)^{-3/2}\inte \frac{e^{i\mu\cdot z}}{\sqrt{c^2|\mu|^2 +m^2c^4}-mc^2+\lam _c}d\mu=\frac{1}{c}\,g_c^{-1}(z), \\
\quad
\end{equation}
where
\[
g_c(\mu )=\frac{1}{\sqrt{ |\mu|^2 +m^2c^2}-mc+\frac{\lam _c}{c}}, \ \ \mu\in \R^3.
\]
In view of (\ref{R:e:6}), we next define
\[
H_c+\frac{\lam _c}{c}:=\sqrt{ -\Delta +m^2c^2}-mc+\frac{\lam _c}{c},
\]
so that
 \begin{equation}\label{R:e:7}
\Big( H_c+\frac{\lam _c}{c}\Big)^{-1}=\int^\infty_0e^{-t\frac{\lam _c}{c}}e^{-tH_c}dt=\int^\infty_0e^{-t(\frac{\lam _c}{c}-mc) }e^{-t\sqrt{ p^2 +m^2c^2}}dt,\quad p=-i\nabla .
\end{equation}
Note from pp. 183 of \cite{LL} that
\[
e^{-t\sqrt{ p^2 +m^2c^2}}(z)=\frac{m^2c^2}{2\pi ^2}\frac{t}{t^2+|z|^2}K_2\big(mc\sqrt{ t^2 +|z|^2}\big),\ \ z\in\R^3,
\]
where $K_2(\cdot)>0$ denotes the modified Bessel function of the third kind.
We then derive from above that
 \begin{equation}\label{R:e:8}
G_c(z)=\frac{1}{c}\,g_c^{-1}(z)=\frac{m^2c}{2\pi ^2}\int^\infty_0e^{-t(\frac{\lam _c}{c}-mc) }\frac{t}{t^2+|z|^2}K_2\big(mc\sqrt{ t^2 +|z|^2}\big)dt.
\end{equation}
Recall from \cite{H} that there exist positive constants $\bar M_1$ and $\bar M_2$, independent of $c>0$, such that
\begin{equation}\label{R:e:9}
K_2(cmw)\le\arraycolsep=1.5pt\left\{\begin{array}{lll}
	 \displaystyle\frac{\bar M_1}{c^2m^2w^2}   \quad   &\mbox{if} & \ \, \displaystyle w<\frac{2}{mc},\\[4mm]
	\displaystyle\frac{\bar M_2e^{-cmw}}{\sqrt{cmw}}   \quad   &\mbox{if}& \,\ \displaystyle w\ge\frac{1}{mc},
\end{array}\right.
\end{equation}
where $w>0$ is a real number. We next follow (\ref{R:e:8}) and (\ref{R:e:9}) to complete the proof by discussing separately the following two cases, which involve very complicated estimates together with some tricks:

\vskip 0.05truein

(1).\ Case of $|z|\ge \frac{1}{mc}$. In this case, we have $\sqrt{t^2+|z|^2}\ge \frac{1}{mc}$ for all $t\ge 0$. We then obtain from  (\ref{R:e:2M}), (\ref{R:e:8}) and (\ref{R:e:9}) that for all sufficiently large $c>0$,
\begin{equation}\label{R:e:10}\arraycolsep=1.5pt\begin{array}{lll}
\displaystyle\frac{2\pi ^2}{\bar M_2m^{\frac{3}{2}}}G_c(z)&\le & \displaystyle\sqrt c\int^\infty_0\frac{t}{(t^2+|z|^2)^{\frac{5}{4}}}e^{-t(\frac{\lam}{c}-mc) -mc\sqrt{ t^2 +|z|^2}} dt\\[3mm]
&=&\displaystyle\int^\infty_{2\sqrt{\frac{m}{\lam}}c^{\frac{3}{2}}|z|} A(t,z)dt+\displaystyle \int^{2\sqrt{\frac{m}{\lam}}c^{\frac{3}{2}}|z|}_{c|z|} A(t,z)dt\\[4mm]
&&+\displaystyle \int^{c|z|}_{\frac{1}{4}\sqrt{\frac{m}{\lam}}c^{\frac{1}{2}}|z|} A(t,z)dt+\displaystyle \int^{\frac{1}{4}\sqrt{\frac{m}{\lam}}c^{\frac{1}{2}}|z|}_0 A(t,z)dt\\[5mm]
&:=&I_A(z)+I_B(z)+I_C(z)+I_D(z),
\end{array}
\end{equation}
where $A(t,z)$ satisfies
\begin{equation}\label{R:e:10A}
A(t,z):=\frac{t\sqrt c}{(t^2+|z|^2)^{\frac{5}{4}}}e^{-t(\frac{\lam}{c}-mc) -mc\sqrt{ t^2 +|z|^2}}.
\end{equation}
For $I_A(z)+I_D(z)$, we note that if $t\ge 0$ satisfies
\[
t\ge 2\sqrt{\frac{m}{\lam}}\,c^{\frac{3}{2}}|z| \ \ \mbox{or}\ \ 0\le t\le \frac{1}{4}\sqrt{\frac{m}{\lam}}\,c^{\frac{1}{2}}|z|,
\]
then one can check that
\begin{equation}\label{R:e:11}
\sqrt{ t^2 +|z|^2}\ge \sqrt{1-\frac{\lam}{mc^2}}\ t+\sqrt{\frac{\lam}{mc}}\,|z|\ge \Big(1-\frac{2\lam}{3mc^2}\Big)t+\sqrt{\frac{\lam}{mc}}\,|z|
\end{equation}
for sufficiently large $c>0$.
We thus obtain from (\ref{R:e:10A}) and (\ref{R:e:11}) that for sufficiently large $c>0$,
\begin{equation}\label{R:e:12}\arraycolsep=1.5pt\begin{array}{lll}
I_A(z)+I_D(z)&\le &\displaystyle\int^\infty_0\frac{t\sqrt c}{(t^2+|z|^2)^{\frac{5}{4}}}e^{-\frac{\lam}{3c}t -\sqrt{ \lam mc}|z|}dt\\[4mm]
&\le &\displaystyle C_1\frac{e^{-\sigma |z|}}{\sqrt c \,|z|}\int^\infty_0\frac{t\sqrt c}{(t^2+|z|^2)^{\frac{5}{4}}}dt\le C_2\frac{e^{-\sigma |z|}}{|z|^\frac{3}{2}},
\end{array}\end{equation}
where $\sigma >0$ is arbitrary, and the constants $C_1>0$ and $C_2>0$ are independent of $c>0$.

For $I_B(z)$, we observe that if $t> 0$ satisfies
\[
c|z|\le t\le 2\sqrt{\frac{m}{\lam}}\,c^{\frac{3}{2}}|z|,
\]
then we have
\begin{equation}\label{R:e:13}
\sqrt{ t^2 +|z|^2}\ge \sqrt{1-\frac{\lam}{mc^2}}\,t+\sqrt{\frac{\lam}{mc^2}}\,|z|\ge \Big(1-\frac{2\lam}{3mc^2}\Big)t+\sqrt{\frac{\lam}{mc^2}}\,|z|
\end{equation}
for sufficiently large $c>0$. We thus obtain from (\ref{R:e:10A}) and (\ref{R:e:13}) that for sufficiently large $c>0$,
\begin{equation}\label{R:e:14}\arraycolsep=1.5pt\begin{array}{lll}
I_B(z)&\le &\displaystyle\int^{2\sqrt{\frac{m}{\lam}}c^{\frac{3}{2}}|z|}_{c|z|}\frac{t\sqrt c}{(t^2+|z|^2)^{\frac{5}{4}}}e^{-\frac{\lam}{3c}t -\sqrt{ \lam m}|z|}dt\\[4mm]
&\le &\displaystyle C_3 e^{-\sqrt{ \lam m}|z|} \int^\infty_{c|z|}\frac{t\sqrt c}{(t^2+|z|^2)^{\frac{5}{4}}}dt\le C_4\frac{e^{-\sqrt{ \lam m}|z|}}{\sqrt{|z|}},
\end{array}\end{equation}
where the constants $C_3>0$ and $C_4>0$ are independent of $c>0$.

As for $I_C(z)$, we get that if $t> 0$ satisfies
\[
C_0c^{\frac{1}{2}}|z|:=\frac{1}{4}\sqrt{\frac{m}{\lam}}\,c^{\frac{1}{2}}|z|\le t\le c|z|,
\]
then we have
\begin{equation}\label{R:e:15}
\sqrt{ t^2 +|z|^2}=t\sqrt{ 1 +\Big(\frac{|z|}{t}\Big)^2}\ge t+\big(\frac{1}{2}-\eps\big)\frac{|z|^2}{t}
\end{equation}
for sufficiently large $c>0$, where $0<\eps<\frac{1}{4}$ is arbitrary. We thus obtain from (\ref{R:e:10A}) and (\ref{R:e:15}) that for sufficiently large $c>0$,
\begin{equation}\label{R:e:16}\arraycolsep=1.5pt\begin{array}{lll}
I_C(z)&\le &\sqrt c\displaystyle\int_{C_0c^{\frac{1}{2}}|z|}^{c|z|}\frac{t}{(t^2+|z|^2)^{\frac{5}{4}}}e^{-\frac{\lam}{c}t -(\frac{1}{2}-\eps)\frac{mc}{t}|z|^2}dt\\[4mm]
&\le &-2\sqrt c\displaystyle \int_{C_0c^{\frac{1}{2}}|z|}^{c|z|}e^{-(\frac{1}{2}-\eps)\frac{mc}{t}|z|^2}dt^{-\frac{1}{2}}\\[4mm]
&:=&2\displaystyle \sqrt{\frac{c}{|z|}}\displaystyle \int^{C_0^{-\frac{1}{2}}c^{-\frac{1}{4}}}_{c^{-\frac{1}{2}}}e^{-(\frac{1}{2}-\eps)mc|z| s^2}ds.
\end{array}\end{equation}
Note that if $\frac{1}{\sqrt c}\le s$, then $2\sqrt cds\le cds^2$. We thus derive from (\ref{R:e:16})  that for $\tau:=s^2>0$,
\begin{equation}\label{R:e:17}
I_C (z)\le  \displaystyle\frac{c}{\sqrt{|z|}} \int^{\frac{1}{c_0\sqrt c}}_{ \frac{1}{c}}e^{-(\frac{1}{2}-\eps)mc|z|\tau}d\tau\le C_5\frac{e^{-(\frac{1}{2}-\eps)m|z|}}{|z|^{\frac{3}{2}}},
\end{equation}
where the constant  $C_5>0$ is also independent of $c>0$, and $0<\eps<\frac{1}{4}$ is arbitrary as before.

Following (\ref{R:e:10}), we now conclude from above that for $0<\delta_0:= \min\{ \big(\frac{1}{2}-\eps\big)m,\sqrt{\lam m}\}$,  where  $0<\eps<\frac{1}{4}$ is arbitrary, there exists a constant $M_0:=M_0(\delta_0)>0$ such that  for all sufficiently large $c>0$,
\begin{equation}\label{R:e:18}
G_c(z) \le   M_0\frac{e^{-\delta _0|z|}}{\min\{|z|^{\frac{1}{2}}, |z|^{\frac{3}{2}}\}},\ \ \mbox{if}\ \ |z|\ge \frac{1}{mc}.
\end{equation}
This further implies that for each $0<\delta_1< \min\{ \frac{m}{2},\sqrt{\lam m}\}$, there exists a constant $M_1:=M_1(\delta_1)>0$ such that for  $|z|\ge \frac{1}{mc}$,
\begin{equation}\label{R:e:18A}
G_c(z) \le   M_1\frac{e^{-\delta _1|z|}}{ |z|^2}
\end{equation}
uniformly for all sufficiently large $c>0$.

\vskip 0.05truein

(2).\ Case of $|z|\le \frac{1}{mc}$. In this case, we deduce from  (\ref{R:e:2M}), (\ref{R:e:8}) and (\ref{R:e:9}) that for all sufficiently large $c>0$,
\begin{equation}\label{R:e:19}\arraycolsep=1.5pt\begin{array}{lll}
\displaystyle 2\pi ^2G_c(z)&\le & \displaystyle\bar M_2m^\frac{3}{2}\sqrt c\int^\infty_{\frac{1}{mc}}\frac{t}{(t^2+|z|^2)^{\frac{5}{4}}}e^{-t(\frac{\lam}{c}-mc) -mc\sqrt{ t^2 +|z|^2}} dt\\[4mm]
&&+\displaystyle \frac{\bar M_1}{c} \int_0^{\frac{1}{mc}}\frac{t}{(t^2+|z|^2)^2}e^{-(\frac{\lam}{c}-mc)t} dt\\[4mm]
&:=&I_1(z)+I_2(z).
\end{array}
\end{equation}
Similar to (\ref{R:e:10}) and (\ref{R:e:18}), one can obtain that for all sufficiently large $c>0$,
\begin{equation}\label{R:e:20}
I_1(z)\le C_6\sqrt c  \int^\infty_0\frac{t}{(t^2+|z|^2)^{\frac{5}{4}}}e^{-t(\frac{\lam}{c}-mc) -mc\sqrt{ t^2 +|z|^2}} dt \le   \frac{M_2}{|z|^2},\ \ \mbox{if}\ \ |z|\le \frac{1}{mc},
\end{equation}
where the constants $C_6>0$ and $M_2>0$ are independent of $c>0$.  As for $I_2(z)$, we infer that
\begin{equation*}
\begin{split}
I_2(z)&\le \frac{C_7}{c} \int_0^{\frac{1}{mc}}\frac{t}{(t^2+|z|^2)^2} dt=\frac{C_7}{2c|z|^2} \int_0^{\frac{1}{mc|z|}}\frac{1}{(s^2+1)^2} ds^2 \\
&= \frac{C_7}{2c|z|^2}\frac{(mc|z|)^2}{1+(mc|z|)^2}\le\frac{C_7}{2c|z|^2},\ \ \mbox{if}\ \ |z|\le \frac{1}{mc},
\end{split}
\end{equation*}
where the constant $C_7>0$ is independent of $c>0$. We therefore derive from (\ref{R:e:19}) and above that for all sufficiently large $c>0$,
\begin{equation}\label{R:e:21}
G_c(z) \le   \frac{M_3}{|z|^2},\ \ \mbox{if}\ \ |z|\le \frac{1}{mc},
\end{equation}
where the constant $M_3>0$ is also independent of $c>0$.

We finally conclude from (\ref{R:e:18A}) and (\ref{R:e:21}) that (\ref{E:e:3}) holds true, and we are done.
\qed

\vskip 0.05truein

\noindent \textbf{Proof of Lemma \ref{lem:2.1}.} We first prove that the positive solution $Q_c>0$ of (\ref{eq2.7}),
where the Lagrange multiplier  $\mu_c$ is as in (\ref{eq1.09C}), satisfies the following exponential decay
 \begin{equation}\label{eF:3}
|Q_c|\le Ce^{-\delta |x|}\ \text{ in}\,\ \R^3
\end{equation}
uniformly for all sufficiently large $c>0$, where the constants $\delta>0$ and $C=C(\delta)>0$ are independent of $c>0$. Actually, recall from (\ref{eq2.7}) that $Q_c$ can be rewritten as
 \begin{equation}\label{R:e:22}
Q_c(x)=-\inte G_c(x-y)V(y)Q_c(y)dy,
\end{equation}
where $G_c(x-y)$ is the Green's function of $\big(\bar H_c+(-\mu _c)\big)^{-1}$ defined by (\ref{R:e:2D}), and the potential $V(x):=-(\frac{1}{|x|}\ast Q_c^2)(x)$ satisfies $V(x)\in C^0(\R^3)$ and $\lim _{|x|\to\infty}V(x)=0$. Since $-\mu_c\to \lam >0$ as $c\to\infty$ in view of (\ref{eq1.09C}), $G_c(x-y)$ satisfies the exponential decay of Lemma \ref{lem:2.3}. Following the above properties, the uniformly exponential decay (\ref{eF:3}) as $c\to\infty$ can be proved in a similar way of \cite[Appendix C]{FJCMP} and \cite[Theorem 2.1]{H}, where the Slaggie-Wichmann method (e.g. \cite{H}) is employed.

To finish the proof of Lemma \ref{lem:2.1}, we next rewrite the solution $\varphi_c\in H^{\frac{1}{2}}(\R^3)$ of (\ref{e:2})
as
\begin{equation}\label{R:e:23}
\varphi_c(x)=\big(\bar H_c+(-\mu _c)\big)^{-1}\big(V_1\varphi_c+2k_1V_2(\varphi_c)+k_2(c)Q_c\big),
\end{equation}
where the operator $\bar H_c$ satisfies (\ref{R:e:2D}) as before and
\begin{equation}\label{R:e:24}
V_1= \frac{1}{|x|}\ast Q_c^2\,,\quad V_2(\varphi_c)=\Big\{\frac{1}{|x|}\ast \big(Q_c \varphi_c\big)\Big\}Q_c.
\end{equation}
Here $\mu_c\in\R$ satisfies (\ref{eq1.09C}), $k_1\ge 0$, and $k_2(c)\in \R$ is bounded uniformly in $c>0$. Note from (\ref{R:e:23}) that $\varphi_c\in H^{\frac{1}{2}}(\R^3)$ solves
\begin{equation}\label{R:e:22}
\varphi_c(x)=\inte G_c(x-y)\big(V_1\varphi_c+k_1V_2(\varphi_c)+k_2(c)Q_c\big)(y)dy,
\end{equation}
where the Green's function $G_c(\cdot)$ of $\big(\bar H_c+(-\mu _c)\big)^{-1}$ satisfies as before the  uniformly exponential decay of Lemma \ref{lem:2.3} in view of (\ref{eq1.09C}). Since $Q_c$ satisfies the uniformly exponential decay (\ref{eF:3}) as $c\to\infty$, the uniformly exponential decay (\ref{e:3}) of $\varphi_c$ as $c\to\infty$ can be further proved in a similar way of \cite[Lemma 4.9]{FJ}. We omit the detailed proof for simplicity. This completes the proof of Lemma \ref{lem:2.1}.
\qed

\section{Proof of Theorem \ref{th2}}

Following the refined estimates of previous section, in this section we shall complete the proof of Theorem \ref{th2}. We begin with the following two lemmas.

\begin{lem}\label{le2.1}
Suppose $Q_\infty$ is the unique radially symmetric positive solution of (\ref{eq1.8}) and let the operator $L_+$ be defined by (\ref{eq1.10}). Then we have
\begin{equation}
L_+(x\cdot \nabla Q_\infty+2Q_\infty)=-2\lambda Q_\infty,
\end{equation}
where $\lam>0$ is as in (\ref{eq1.8}).
\end{lem}

\noindent \textbf{Proof.} Direct calculations give that
\begin{equation*}
\begin{split}
-\Delta(x\cdot \nabla Q_\infty)&=-\sum_{i=1}^3\partial_{i}\big(\partial_{i}(\sum_{j=1}^3x_j \partial_{j}Q_\infty)\big)
=-\sum_{i,j=1}^3\partial_i\big(\delta_{ij}\partial_j Q_\infty+x_j\partial_{ij}Q_\infty\big)\\
&=-\sum_{i,j=1}^3\big(2\delta_{ij}\partial_{ij }Q_\infty+x_j\partial_{iij}Q_\infty\big)=-2\Delta Q_\infty-x\cdot\nabla(\Delta Q_\infty),
\end{split}
\end{equation*}
and
\begin{equation*}
2Q_\infty\big[|x|^{-1}\ast (Q_\infty x\cdot\nabla Q_\infty)\big]=Q_\infty\big[|x|^{-1}\ast (x\cdot\nabla Q_\infty^2)\big].
\end{equation*}
We then have
\begin{equation}\label{eq2.2}
\begin{split}
L_+(x\cdot \nabla Q_\infty)&=-\frac{1}{m}\Delta Q_\infty-\frac{1}{2m}x\cdot\nabla\big(\Delta Q_\infty\big)+\lambda x\cdot \nabla Q_\infty\\
&-\big(|x|^{-1}\ast Q_\infty^2)(x\cdot \nabla Q_\infty)-Q_\infty\big[|x|^{-1}\ast (x\cdot\nabla Q_\infty^2)\big].
\end{split}
\end{equation}
Taking the action $x\cdot \nabla $ on  (\ref{eq1.8}),  we deduce  that
$$-\frac{1}{2m}x\cdot \nabla(\Delta Q_\infty)+\lambda x\cdot \nabla Q_\infty=\big(|x|^{-1}\ast Q_\infty^2\big)\big(x\cdot \nabla Q_\infty\big)+Q_\infty x\cdot\big(|x|^{-1}\ast \nabla Q_\infty^2\big).$$
Together with (\ref{eq2.2}), this indicates that
\begin{equation}\label{eq2.3}
\begin{split}
L_+(x\cdot \nabla Q_\infty)&=-\frac{1}{m}\Delta Q_\infty+Q_\infty x\cdot\big(|x|^{-1}\ast \nabla Q_\infty^2\big)-Q_\infty\big[|x|^{-1}\ast (x\cdot\nabla Q_\infty^2)\big].
\end{split}
\end{equation}
Since
\begin{equation}\label{eq2.33}
\begin{split}
& x\cdot\big(|x|^{-1}\ast \nabla Q_\infty^2\big)-|x|^{-1}\ast (x\cdot\nabla Q_\infty^2)\\
  =&\inte \frac{x\cdot\nabla Q_\infty^2(y)}{|x-y|}dy-\inte\frac{y\cdot\nabla Q_\infty^2(y)}{|x-y|}dy \\
 =&\inte\frac{(x-y)\cdot\nabla Q_\infty^2(y)}{|x-y|}dy=-\sum_{i=1}^3\inte\partial_{y_i}\Big(\frac{x-y}{|x-y|}\Big) Q_\infty^2(y)dy \\
 =&2\inte\frac{Q_\infty^2(y)}{|x-y|}dy=2|x|^{-1}\ast  Q_\infty^2,
\end{split}
\end{equation}
it follows from (\ref{eq2.3}) that
\begin{equation}\label{eq2.4}
\begin{split}
L_+(x\cdot \nabla Q_\infty)&=-\frac{1}{m}\Delta Q_\infty+2\big(|x|^{-1}\ast  Q_\infty^2\big)Q_\infty.
\end{split}
\end{equation}
Moreover, recall from (\ref{eq1.8}) that
 \begin{equation}\label{eq2.5}
  L_+ Q_\infty=\Big(-\frac{1}{2m}\Delta+\lambda-3|x|^{-1}\ast |Q_\infty|^2\Big)Q_\infty=-2(|x|^{-1}\ast  Q_\infty^2)Q_\infty.
  \end{equation}
Combining (\ref{eq2.4}) with (\ref{eq2.5})  thus yields  that
\begin{equation*}
L_+(x\cdot \nabla Q_\infty+2Q_\infty)=-\frac{1}{m}\Delta Q_\infty-2\big(|x|^{-1}\ast  Q_\infty^2\big)Q_\infty=-2\lambda Q_\infty,
\end{equation*}
and the proof of this lemma is therefore complete.
\qed


\begin{lem}\label{le2.2}
Let $Q_c$ be a radially symmetric positive  minimizer of $\bar e(c)$ defined in (\ref{def:ec}). Then we have the following Pohozaev identity
\begin{equation}\label{eq2.07}
-m^2c^4 \big\langle(-c^2\Delta+m^2c^4)^{-\frac{1}{2}}Q_c,Q_c\big\rangle  +mc^2\inte|Q_c|^2dx+\bar e(c)=0.
\end{equation}
\end{lem}

\noindent \textbf{Proof.}
In the proof of this lemma, we denote $Q_c$ by $Q$ for convenience. We first note that
$$x\cdot \nabla Q(x) =\frac{d}{d\lambda}Q_\lambda(x)\big|_{\lambda=1},\ \text{ where }Q_\lambda(x):=Q(\lambda x).$$
Multiplying $x\cdot \nabla Q(x)$ on both sides of (\ref{eq2.7}) and integrating over $\R^3$, we have
\begin{equation}\label{eq2.8}
\begin{split}
& \big\langle\sqrt{-c^2\Delta+m^2c^4}Q,x\cdot \nabla Q\big\rangle  \\
=&\frac{d}{d\lambda} \big\langle\sqrt{-c^2\Delta+m^2c^4}Q,Q_\lambda\big\rangle \Big|_{\lambda=1}  \\
 =&\frac{d}{d\lambda}  \big\langle(-c^2\Delta+m^2c^4)^\frac{1}{4}Q,(-c^2\Delta+m^2c^4)^\frac{1}{4}Q_\lambda\big\rangle \Big|_{\lambda=1}  \\
 \quad\qquad \overset{\sqrt\lambda x=x'}   = &  \frac{d}{d\lambda} \lambda^{-1} \big\langle(-c^2\Delta+\frac{m^2c^4}{\lambda})^\frac{1}{4}Q\big(\frac{x}{\sqrt\lambda}\big),
 \big(-c^2\Delta+\frac{m^2c^4}{\lambda}\big)^\frac{1}{4}Q(\sqrt\lambda x)\big\rangle \Big|_{\lambda=1} \\
 =&-\inte\big|(-c^2\Delta+m^2c^4)^\frac{1}{4}Q\big|^2dx\\
 \quad &-\frac{m^2c^4}{2} \big\langle(-c^2\Delta+m^2c^4)^{-\frac{3}{4}}Q,(-c^2\Delta+m^2c^4)^\frac{1}{4}Q\big\rangle   \\
 =&- \big\langle\sqrt{-c^2\Delta+m^2c^4}Q,Q\big\rangle -\frac{m^2c^4}{2} \big\langle(-c^2\Delta+m^2c^4)^{-\frac{1}{2}}Q,Q\big\rangle  .
\end{split}
\end{equation}
Moreover, we derive from the exponential decay (\ref{eq1.09}) that
\begin{align}
&\inte (|x|^{-1}\ast Q^2)Q(x\cdot \nabla Q)dx=\frac12\inte (|x|^{-1}\ast Q^2)(x\cdot \nabla Q^2)dx\nonumber\\
&=-\frac{3}{2}\inte(|x|^{-1}\ast Q^2)Q^2dx-\frac{1}{2}\inte\Big[(|x|^{-1}\ast \nabla Q^2)\cdot x\Big] Q^2dx\nonumber\\
&=-\frac{3}{2}\inte(|x|^{-1}\ast Q^2)Q^2dx-\frac{1}{2}\Big[2\inte (|x|^{-1}\ast Q^2) Q^2dx\nonumber\\
&\quad\quad\quad\quad\quad\quad\quad\quad\quad\quad\quad\quad+\iint_{\R^3} \frac{y\cdot\nabla Q^2(y)}{|x-y|}Q^2(x)dydx\Big],\label{eq2.10}
\end{align}
where the argument of deriving (\ref{eq2.33}) is used in the last equality.
Since
\[\begin{split}
\iint_{\R^3} \frac{y\cdot\nabla Q^2(y)}{|x-y|}Q^2(x)dydx =\inte (|x|^{-1}\ast Q^2)(x\cdot \nabla Q^2)dx,
\end{split}\]
we obtain from (\ref{eq2.10}) that
\begin{equation}\label{eq2.11}
\inte (|x|^{-1}\ast Q^2)Q(x\cdot \nabla Q)dx=-\frac54\inte(|x|^{-1}\ast Q^2)Q^2dx.
\end{equation}
One can easily check that
$$\inte Q(x\cdot \nabla Q)dx=-\frac{3}{2}\inte Q^2dx.$$
Thus, it follows from (\ref{eq2.7}), (\ref{eq2.8}) and (\ref{eq2.11}) that
\begin{equation}
\begin{split}
&- \big\langle\sqrt{-c^2\Delta+m^2c^4}Q,Q\big\rangle -\frac{m^2c^4}{2} \big\langle(-c^2\Delta+m^2c^4)^{-\frac{1}{2}}Q,Q\big\rangle  \\
&=-\frac32(mc^2+\mu _c)\inte Q^2dx-\frac54\inte(|x|^{-1}\ast Q^2)Q^2dx.
\end{split}\label{eq2.12}
\end{equation}
By (\ref{eq2.12}), multiplying $Q$ on both sides of  (\ref{eq2.7})  and integrating  over $\R^3$ yield that
\begin{equation}\label{eq2.13}
\begin{split}
&-{m^2c^4} \big\langle(-c^2\Delta+m^2c^4)^{-\frac{1}{2}}Q,Q\big\rangle   \\
&+(mc^2+\mu_c)\inte|Q|^2dx+\frac12\inte(|x|^{-1}\ast Q^2)Q^2dx=0.
\end{split}
\end{equation}
We also derive from (\ref{eq2.7}) that
\begin{equation*}
\begin{split}\mu_c\inte|Q|^2dx&= \big\langle\sqrt{-c^2\Delta+m^2c^4}Q,Q\big\rangle  -mc^2\inte|Q|^2dx-\inte(|x|^{-1}\ast Q^2)Q^2dx\\
&=\bar e(c)-\frac{1}{2}\inte(|x|^{-1}\ast Q^2)Q^2dx,
\end{split}\end{equation*}
which therefore implies that (\ref{eq2.07}) holds true by applying (\ref{eq2.13}).
\qed

\vskip.05truein

Following previous estimates, we are now ready to finish the proof of Theorem \ref{th2}.

\vskip.05truein

\noindent\textbf{Proof of Theorem \ref{th2}.} Up to the phase and translation, it suffices to prove that $\bar e(c)$ in (\ref{def:ec}) admits a unique positive minimizer for sufficiently large $c> 0$. On the contrary, suppose that $Q_{1c}$ and $Q_{2c}$ are two different radially symmetric (about the origin) positive  minimizers of problem (\ref{def:ec}), where $m>0$ is fixed.

Then $Q_{ic}\in H^s(\R^3)$, where $s\ge \frac{1}{2}$, satisfies the following equation
\begin{equation}\label{eq2.14}
\big(\sqrt{-c^2\Delta +m^2c^4}-mc^2\big)Q_{ic}-(|x|^{-1}\ast Q_{ic}^2)Q_{ic}=\mu_{ic} Q_{ic} \ \text{ in }\, \R^3, \ i=1,2,
\end{equation}
where $\mu_{ic} \in\R $ is the  Lagrange multiplier associated to $Q_{ic}$ for $i=1,2$. Since $Q_{1c}\not\equiv Q_{2c}$ in $\R^3$, we define
\begin{equation}\label{eq2.15}
w_c(x):=\frac{Q_{1c}(x)-Q_{2c}(x)}{\|Q_{1c}-Q_{2c}\|_{L^\infty(\R^3)}} \ \text{ in }\, \R^3.
\end{equation}
It then follows from (\ref{eq2.14}) that
\begin{equation}\label{eq2.16}
\begin{split}
&\big(\sqrt{-c^2\Delta +m^2c^4}-mc^2\big)w_{c}-\big(|x|^{-1}\ast Q_{1c}^2\big)w_{c}-\Big\{|x|^{-1}\ast \big[(Q_{1c}+Q_{2c})w_c\big]\Big\}Q_{2c}\\
&=\mu_{2c} w_c+\frac{\mu_{1c}-\mu_{2c}}{\|Q_{1c}-Q_{2c}\|_{L^\infty(\R^3)}}Q_{1c} \ \text{ in }\, \R^3.
\end{split}
\end{equation}
Recall from (\ref{eq1.09C}) that
\begin{equation}\label{eq2.27}
\lim_{c\to\infty}\mu_{ic}=-\lambda<0, \ i=1,2.
\end{equation}
We also note from (\ref{eq2.14}) that
\begin{equation*}\label{eq2.17}
\mu_{ic}=\bar e(c)-\frac{1}{2}\inte \big(|x|^{-1}\ast Q_{ic}^2\big)Q_{ic}^2dx, \ i=1,2,
 \end{equation*}
which implies that
\begin{equation}\label{eq2.17:K}
\begin{split}
 \frac{\mu_{1c}-\mu_{2c}}{\|Q_{1c}-Q_{2c}\|_{L^\infty(\R^3)}} = -\frac{1}{2}\inte &\Big\{ (|x|^{-1}\ast Q_{1c}^2)(Q_{1c}+Q_{2c})w_{c}\\
  &+\Big[|x|^{-1}\ast \big((Q_{1c}+Q_{2c})w_c\big)\Big]Q_{2c}^2\Big\}dx:=k_2(c)\in\R,
\end{split}
 \end{equation}
where $k_2(c)$ is bounded uniformly in $c>0$ by (\ref{eq1.09}) and (\ref{B:M}).
Applying Lemma \ref{lem:2.1} to the equation (\ref{eq2.16}), we then deduce from (\ref{eq2.27}) and (\ref{eq2.17:K}) that there exist $\delta>0$ and $C=C(\delta)>0$, which are independent of $c>0$, such that
 \begin{equation}\label{e:3:3}
|w_c|\le Ce^{-\delta |x|}\ \text{ in}\,\ \R^3
\end{equation}
uniformly as $c\to\infty$. Similar to the proof of Lemma \ref{lem:2.1}, following (\ref{eq2.16}) to consider the equation of $\frac{\partial w_c}{\partial x_i}$ ($i=1, 2, 3$), we further derive from (\ref{eq2.27})--(\ref{e:3:3}) that
there exist $\delta _1>0$ and $C_1=C_1(\delta_1)>0$, which are independent of $c>0$, such that
 \begin{equation}\label{e:3:3B}
|\nabla w_c|\le C_1e^{-\delta _1|x|}\ \text{ in}\,\ \R^3
\end{equation}
uniformly as $c\to\infty$.

Rewrite the equation (\ref{eq2.16}) as
\begin{equation}\label{B:eq2.16}
\begin{split}
\big(\bar{H}_c+\mu\big)w_{c}=&\Big\{\mu+\mu_{2c} +\big(|x|^{-1}\ast Q_{1c}^2\big)\Big\}w_c\\
&+k_2(c)Q_{1c} +\Big\{|x|^{-1}\ast \big[(Q_{1c}+Q_{2c})w_c\big]\Big\}Q_{2c}\ \text{ in }\, \R^3
\end{split}
\end{equation}
for any $\mu\in \R$, where  the uniformly bounded function $k_2(c)\in\R$ is as in (\ref{eq2.17:K}), and the operator $\bar {H}_c$ satisfies
\[
\bar{H}_c:=\sqrt{-c^2\Delta +m^2c^4}-mc^2.
\]
Since $\|w _c\|_{L^\infty(\R^3)}\le 1$ for all $c>0,$ we obtain from (\ref{eq1.09}), (\ref{B:M}) and (\ref{2A:E1}) that for sufficiently large $c>0$,
\begin{equation}\label{C:eq2.16}
 \big\| |x|^{-1}\ast Q_{1c}^2 \big\| _{L^\infty(\R^3)}<M, \ \   \big\|  |x|^{-1}\ast \big[(Q_{1c}+Q_{2c})w_c\big]  \big\|  _{L^\infty(\R^3)}<M,
\end{equation}
where $M>0$ is independent of $c>0$. In view of (\ref{B:M}) and (\ref{C:eq2.16}), the similar argument of \cite[Theorem 4.1(i)]{FJCMP}  or \cite[Proposition 4.2]{Choi} applied to (\ref{B:eq2.16}) and (\ref{eq2.27}) yields that
\begin{equation}\label{F:M}
w_{c}\in H^s(\R^3) \ \   \mbox{for all}\, \ s\ge \frac{1}{2},
\end{equation}
which further implies the uniform smoothness of $w_c$ in $c>0$.

We next rewrite the equation (\ref{eq2.16}) as
\begin{equation}\label{eq2.16K}
\begin{split}
& \Big(-\frac{1}{2m}\Delta -\mu_{2c}\Big)w_{c} -\big(|x|^{-1}\ast Q_{1c}^2\big)w_{c}-\Big\{|x|^{-1}\ast \big[(Q_{1c}+Q_{2c})w_c\big]\Big\}Q_{2c}\\
&=k_2(c)Q_{1c}- F_{c}(\nabla) w_c \ \text{ in }\, \R^3,
\end{split}
\end{equation}
where the uniformly bounded function $k_2(c)\in\R$ is again as in (\ref{eq2.17:K}), and the pseudo-differential operator $F_{c}(\nabla)$ is the same as (\ref{2D:2}) with the symbol
$$
F_{c}(\xi):=\sqrt{c^2|\xi|^2 +m^2c^4}-mc^2-\frac{ |\xi|^2}{2m}, \ \  \xi\in\R^3.$$
Note that $F_{c}(\nabla)$ satisfies the estimate (\ref{2D:2D}). We then derive from (\ref{2D:2D}) and (\ref{F:M}) that
\begin{equation}\label{2K:2D}
\|F_{c} (\nabla)w_{c}\|_{H^s(\R^3)}\le \frac{M_1}{m^3c^2}\|w_c\|_{H^{s+4}(\R^3)}<\frac{M_2}{m^3c^2} \ \ \mbox{for all}\ \ s\ge \frac{1}{2},
\end{equation}
where $M_1>0$ and $M_2>0$ are independent of $c>0$.

Using the uniformly exponential decays (\ref{eq1.09}) and (\ref{e:3:3}),  by the standard elliptic regularity we derive from (\ref{B:M}), (\ref{eq2.16K}) and (\ref{2K:2D}) that $\|w _c\|_{C^{\alpha }_{loc}(\R^3)}\le M_3$ for some $\alp \in (0,1)$, where the constant $M_3>0$ is independent of $c$. Therefore, there exists a function $w _0=w _0(x)$ such that up to a subsequence if necessary, we have
$$w_c\to w_0\ \text{ in }\ C_{\rm loc}(\R^3)\ \text{ as }\ c\to\infty.$$
Moreover, applying the estimates (\ref{eq2.27}), (\ref{eq2.17:K}) and (\ref{2K:2D}), we deduce from (\ref{eq2.16K}) that $w_0$ satisfies
\begin{equation}\label{eq2.18A}
L_+ w_0
=-Q_\infty\inte\Big\{(|x|^{-1}\ast Q_{\infty}^2)Q_\infty w_{0}+\big[|x|^{-1}\ast (Q_{\infty}w_0)\big]Q_{\infty}^2\Big\}dx,
\end{equation}
where the uniformly exponential decay (\ref{eq1.09}) is also used. Applying (\ref{eq1.12}) and Lemma 3.1, we now derive from (\ref{eq2.18A}) that there exist constants $b_0$ and $c_i$ ($i=1, 2, 3$) such that  \[
w_0=b_0\big(x\cdot \nabla Q_\infty+2Q_\infty\big)+\sum ^3_{i=1}c_i\frac{\partial Q_\infty}{\partial x_i}.
\]
Further, since $Q_{1c}$ and $Q_{2c}$ are both radially symmetric in $|x|$ for all $c>0$, the definition of $w_c$ implies that $w_0$ is also radially symmetric in $|x|$, i.e., $w_0\in L^2_{rad}(\R^3)$. Applying \cite[Proposition 2]{L09}, it then follows from the above expression that
\begin{equation}\label{eq2.18}
w_0=b_0\big(x\cdot \nabla Q_\infty+2Q_\infty\big)\ \text{ for some }\, b_0\in \R.
\end{equation}


We next prove $b_0=0$ in (\ref{eq2.18}) so that $w_0\equiv 0$ in $\R^3$. Indeed, multiplying (\ref{eq2.07}) by $Q_{1c}$ and $Q_{2c}$ respectively,  and integrating over $\R^3$, we obtain that
\begin{equation}\label{eq2.19}
\begin{split}
-mc^2  \big\langle(-c^2\Delta+m^2c^4)^{-\frac{1}{2}}w_c,Q_{1c}\big\rangle&-mc^2  \big\langle(-c^2\Delta+m^2c^4)^{-\frac{1}{2}}Q_{2c},w_c\big\rangle  \\
&+\inte w_c(Q_{1c}+Q_{2c})dx=0.
\end{split}
\end{equation}
Moreover, since  $\frac{1}{\sqrt{1+t}}=1-\frac{t}{2}+O(t^2)$ as $t\to0$ and $w_c$ is smooth for all $c>0$,  we have
\[\begin{split}
mc^2\big(-c^2\Delta+m^2c^4\big)^{-\frac{1}{2}}=  \Big(1-\frac{\Delta}{m^2c^2} \Big)^{-\frac{1}{2}}=\Big[1+\frac{\Delta}{2m^2c^2}+O(\frac{1}{m^4c^4})\Big]\ \ \mbox{as}\ \, c\to\infty.
\end{split}\]
Putting it into (\ref{eq2.19}) yields that
\begin{equation*}
-\frac{1}{2m}\inte \big(Q_{1c}\Delta w_c +w_c\Delta Q_{2c}\big)dx+O\Big(\frac{1}{m^3c^2}\Big)=0\ \ \mbox{as}\ \, c\to\infty,
\end{equation*}
where the exponential decays (\ref{eq1.09}), (\ref{e:3:3}) and (\ref{e:3:3B}) are used again.
Applying (\ref{eq1.09}), (\ref{eq1.09B}) and (\ref{F:M}), it then follows from above that
\begin{equation}\label{eq2.31}
\inte \nabla w_0\nabla Q_\infty dx =0.
\end{equation}
We thus conclude from (\ref{eq2.18}) and (\ref{eq2.31}) that
\begin{equation}
0=b_0\inte \nabla \big(x\cdot \nabla Q_\infty+2Q_\infty\big)\nabla Q_\infty dx=\frac{3}{2}b_0\inte |\nabla Q_\infty|^2 dx,
\end{equation}
which therefore implies that $b_0=0$ in (\ref{eq2.18}) and thus $w_0\equiv 0$ in $\R^3$.

We are now ready to derive a contradiction. In fact, let $y_c\in \R^3$ be a point satisfying $|w_c(y_c)|=\|w_c \|_{L^\infty(\R^3)}=1$. Since it follows from (\ref{e:3:3}) that $w_c$ admits the exponential decay uniformly for all $c>0$, we have $|y_c|\le M$ uniformly in $c$ for some constant $M>0$. Therefore, we obtain that $w_c\to w_0\not\equiv 0$ uniformly on $\R^3$ as $c\to\infty$, which however contradicts to the fact that $w_0 \equiv 0$ on $\R^3$. This completes the proof of Theorem \ref{th2}.
\qed
\vspace {.5cm}

\noindent {\bf Acknowledgements:}  The authors thank Professor Enno Lenzmann very much for his helpful discussions on the subject of the present work.


\end{document}